***************************************************************************
\documentstyle[12pt]{article}
\def\x{{\bf x}}
\def\r{{\bf r}}
\def\f{{\bf f}}
\def\q{{\bf q}}
\def\p{{\bf p}}
\def\v{{\bf v}}
\def\u{{\bf u}}
\def\begineq{\begin{equation}}
\def\endeq{\end{equation}}
\begin{document}
\title{Scale resolved intermittency in turbulence}
\author{Siegfried Grossmann $^1$ and Detlef Lohse $^{1,2}$}
\maketitle

\bigskip

\begin{tabular}{ll}

$^1$ & Fachbereich Physik, Philipps-Universit\"at,\\
& Renthof 6, D-35032 Marburg, Germany \\\\
$^2$ & The James Franck Institute, The University of Chicago,\\
& 5640 South Ellis Avenue, Chicago, IL 60637, USA\\\\
\end{tabular}

\date{}
\maketitle
\bigskip
\bigskip

The deviations $\delta\zeta_m$ ("intermittency corrections") from classical
("K41")  scaling   $\zeta_m=m/3$ of the $m^{th}$ moments of  the  velocity
differences  in high Reynolds number turbulence  are  calculated,
extending  a  method  to approximately  solve  the  Navier-Stokes
equation described earlier. We suggest to introduce the notion of
scale resolved intermittency corrections $\delta\zeta_m(p)$, because we find
that these $\delta\zeta_m(p)$ are large in the viscous subrange, moderate in
the nonuniversal stirring subrange but,  surprisingly,  extremely
small if not zero in the inertial subrange. If ISR
intermittency corrections persisted
in experiment up to the large  Reynolds number
limit, our calculation would show, that this could be due to the
opening of phase space for larger wave
vectors. In the higher order velocity moment $\langle|u(p)|^m\rangle$
the crossover  between  inertial and viscous subrange
is $(10\eta  m/2)^{-1}$,  thus the inertial subrange is  {\it smaller}  for
higher moments.
\\
\vspace{1cm}\noindent

\newpage
Experimentally  turbulent  flow has long been  known  to  be
intermittent \cite{bat49}.  A signal is called intermittent,  if there are
relatively  calm  periods which are  irregularly  interrupted  by
strong  turbulent bursts either in time or in  space.
Correspondingly,  the probability density function (PDF) develops enhanced
tails  of  large  fluctuations  and a  center  peak  due  to  the
abundance  of calm periods,  i.e.,  the PDF becomes of  stretched
exponential type instead of being Gaussian.  This also means that
the r-scaling exponents $\zeta_m$ of the velocity differences,
\begineq
<|\v_r (\x )|^m>
\equiv  <|{\bf u} ({\bf x} + {\bf r},t) - {\bf u} ({\bf x},t)|^m>
\propto r^{\zeta_m},
\endeq
do  not vary linearly with m,  namely as m/3,  as was  originally
suggested  by  dimensional analysis of  the  universal,  inertial
subrange of fully developed turbulent flow \cite{kol41,wei48}.
Any deviations
$\delta\zeta_m=\zeta_m-m/3$ are called intermittency corrections.

{\it Phenomenological} intermittency {\it models} describe the  measured
intermittency  corrections $\delta\zeta_m$ more or less successfully.  For  a
detailed discussion see e.g. \cite{men91b,gnlo92b}.
But from our point of view an
understanding  of  intermittency  has to come  from  the  {\it
Navier-Stokes equation}.

As  full  simulations  for  high  Reynolds  number  ($Re\approx 10^6$)
turbulence are out of range even for near future  computers,  one
is thrown on approximations of the Navier-Stokes dynamics.

The  main idea of such an approximation has been  introduced
by  us  in \cite{gnlo92b,egg91a}.  Meanwhile we have  considerably  improved
our
approach  and in this paper we employ it to determine the
intermittency corrections
$\delta\zeta_m$. For completeness, we briefly repeat how
our approximation scheme works.

It  starts from the common Fourier series in terms of  plane
waves $\exp{(i\p\cdot\x)}$, $\p=(p_i)$, $p_i=n_i L^{-1}$,
$n_i=0,\pm 1,\pm 2,\dots$. The periodicity
volume is $(2\pi L)^3$,  L is the outer length scale.  To deal feasibly
with the many scales present in turbulent flow,  we only admit  a
geometrically scaling subset K of wave vectors in the Fourier sum,
$K=\cup_l  K_l$,  thus
$u_i(\x,t)=\sum_{\p\in K}  u_i(\p,t) \exp{(i\p\cdot\x )}$.
Therefore we have called our approximation scheme ``Fourier-Weierstrass
decomposition'' \cite{gnlo92b,egg91a}.
$K_0=\{\p_n^{(0)}$,  $n=1,...,N\}$ contains
appropriately chosen wave vectors, which already have quite
different lengths but dynamically interact to a good degree.  The
$K_l=\{\p_n^{(l)}=2^l \p_n^{(0)},\quad n=1,...,N\}$, $l=1,...,l_{max}$,
are scaled replica of $K_0$
which represent smaller and smaller eddies.  $l_{max}$ is chosen large
enough  to  guarantee  that the amplitudes  $\u(\p_n^{l_{max}}    ,t)$
of  the smallest eddies are practically zero.  Of course, $l_{max}$ depends on
the viscosity $\nu$ and thus on Re.

We solve the Navier-Stokes equation for incompressible  flow
(i.e., $\p\cdot\u(\p)=0$) in the subspace defined by the wave vector set K,
\begin{eqnarray}
\dot u_i(\p)  & = &
-iM_{ijk}(\p) \sum_{\q_1,\q_2 \in K, \q_1+\q_2=\p}
 u_j({\bf q_1})
u_k({\bf q_2})
-\nu {\bf p}^2 u_i({\bf p}) + f_i(\p).
\end{eqnarray}
The set $K_0$ is chosen in a way that as many triadic  Navier-Stokes
interactions $\p=\q_1+\q_2$ as possible are admitted.  The degree of the
nonlocality in p-space of any triadic interaction can be
characterized  by the quantity $s:=max(p,q_1,q_2)/min(p,q_1,q_2)$.  We  allow
for s up to 5.74. To force the flow permanently we choose
\begin{eqnarray}
\f (\p,t) & = & {\epsilon \
\u(\p ,t) \over \sum_{\q \in K_{in}}
 |\u (\q ,t)|^2}
\qquad \hbox{for} \quad \p \in K_{in}, \\
\f (\p,t) & = & 0
\qquad \hbox{for} \quad \p \notin K_{in} \nonumber,
\end{eqnarray}
as a deterministic,  non-stochastic driving. $K_{in}\subset K_0$ only contains
the wave vectors with the three smallest lengths.  The
corresponding amplitudes $\u(\p,t)$ carry the largest energy.

For  the  numerical calculations the times are  measured  in
units  of  the largest eddies turnover time $L^{2/3}\epsilon^{ -1/3}$
and  the
velocities in units of $(L\epsilon )^{1/3}$.  The Reynolds number can then  be
defined  by $Re=\nu^{ -1}$.  The coupled set of $(3\cdot (l_{max}+1)\cdot N)$
equations
(2)  is  integrated with the Burlirsch-Stoer  integration  scheme
with adaptive stepsize.  All averages are time averages,  denoted
by $\langle \dots\rangle$.

We remark that the density of the admitted wave vectors  per
p-interval decreases as $1/p$ in our reduced waveset approximation,
whereas it increases as $p^2$ in full grid simulations,  see  Fig.1.
But  this shortcoming at the same time is the main  advantage  of
our approximation,  because many more scales than in full
simulations  can be taken into account.  In \cite{egg91a} and \cite{gnlo93g}
we achieved  $Re=2\cdot 10^6$,
i.e.,  3  decades. We used $N=26$,  $l_{max}=10$ and
$N=80$, $l_{max}=12$, respectively.  The main features  of  fully
developed  turbulence  as  chaotic  signals,  scaling,  turbulent
diffusion,  etc.,  are  well described within  our  approximation
\cite{gnlo92b,egg91a}.  In particular,  our solutions show {\it small scale
intermittency}.  This  is  accounted for by a competition  effect  between
turbulent energy transfer downscale and viscous dissipation \cite{gnlo92b}.

The  main improvements of our new treatment in  the  present
paper  are:  (i)  The  number of  modes  per  $K_l$ is  considerably
increased up to N=86 instead of N=38 in \cite{gnlo92b} or N=26 in
\cite{egg91a}.
Thus the number of contributing triades in eq.(2) is much larger,  see
table 1.  (ii) We now also allow for {\it nonlocal} interactions in
p-space.  The  nonlocality  of the waveset K with  respect  to  the
Navier-Stokes dynamics (2) can be quantified by $s_{max}$,  defined as
the maximum of the s-values of all contributing triades.  In  our
former calculations \cite{gnlo92b,egg91a} we had $s_{max} \le 2$,  which means
that eddies
can  at most decay in half size eddies,  whereas now  for  $s_{max}\approx 6$
(see  table 1) sweeping of small eddies on larger ones (up  to  a
factor  of  6)  is  possible.  (iii)  We  can  now  consider  the
individual $\u(\p,t)$ instead of the whole shells $\u^{(l)}=\sum_{\p\in K_l}
\u(\p)$, what
had to be done in \cite{gnlo92b,egg91a}
because of the smaller number of  triadic
interactions.   (iv)  We  are  now  much  beyond  a  shell  model
\cite{gnlo92b,egg91a,sig77},
since on the $|\p|$-axis the elements of the wave vector
subsets $K_l$ interpenetrate and intermingle considerably.

We now offer our results.

The  spectra  $\langle |\u(\p)|^2\rangle$ and
$\langle |\u(\p)|^6\rangle$ calculated  with  N=86
wave vectors  in $K_0$ and with $Re=125~000$ are shown  in  Fig.2a.  For
comparison  the same spectra are also given for N=38 as  used  in
our previous work (Fig.2b,  \cite{gnlo92b}). As expected the scatter becomes
less  with  increasing  N.  We fit the  spectra  with  the  three
parameter functions
\begineq
\langle |\u (\p )|^m\rangle = c_m p^{-\zeta_m} \exp{(-p/p_{D,m})}.
\endeq
In Table 2 the fit parameters  $\zeta_m$ and $p_{D,m}$ are listed.  The ansatz
(4) is theoretically known \cite{foi90} to hold for m=2.  We find that it
also holds for $m>2$ with $p_{D,m}=2p_{D,2}/m$ as one can expect, if in the
VSR  the  higher moments factorize.  For the  dissipative  cutoff
$p_D:=p_{D,2}$ we obtain $p_D=(11\eta )^{-1}$,  where
$\eta =(\nu^3/\epsilon)^{1/4}$ is the Kolmogorov
length.  This well agrees with the long known (experimentally
\cite{my75}  and  theoretically \cite{eff87,gnlo93a})
crossover  between  the  viscous
subrange  VSR  and  the inertial subrange ISR  in  the  structure
function  $D^{(2)}(r)=\langle |\u (\x +\r )-\u (\x )|^2\rangle$
at r  about  $10\eta$ .
According  to  our  finding $p_{D,m}=2p_{D,2}/m$ the  crossover  $p_{D,m}$  in
higher  order
moments
$\langle|u(p)|^{m}\rangle$
occurs  at  smaller p,
namely,  approximately at
${(}10\eta m/2{)}^{-1}$.
The ISR for higher moments is thus definitely smaller.
This does {\it not}
necessarily mean that the ISR for
higher order {\it structure} functions $D^{m}(r)$
is also smaller, because they are not
simply connected with $\langle|u(p)|^{m}\rangle$
via a Fourier transform as in the case $m=2$.
In passing by we remark that by properly
renorming the wave vector p and the spectral
intensity, the spectra (4) can be shown to the {\it universal}
for all Reynolds numbers both in
experiment \cite{she93b} and in full simulations \cite{she93} and in our
approximate Navier Stokes solution \cite{gnlo93g}.

The  intermittency corrections $\delta\zeta_m$ from our overall fit (4)  are
much smaller than the experimental ones around. At the other hand
we observe here as in \cite{gnlo92b}
that there {\it is} much intermittency in the
signals,  at least for small scales.  We therefore determined the
exponents $\zeta_m$  in (4) by fitting  restricted  p-ranges  only.  We
suggest to  introduce "local"  $\zeta_m(p)$. These are defined by local
fits of  the type  (4), using for each wave vector $\p$ the moments
in the local p-decades $[p/\sqrt{10},p\sqrt{10}]$. The cutoff wave vectors are
kept fixed  at their global values $p_D=(11\eta)^{-1}$, $p_{D,m}=2p_D/m$. Also,
as before  (see caption  of table 2) we devide the local  $\zeta_m(p)$
by $\zeta_3(p)$.\footnote{Instead of  defining
$\delta\zeta_m(p)=\zeta_m(p)/\zeta_3(p)-m/3$  one could also
take the  deviation $\tilde\delta\zeta_m(p)$  of  the  $\zeta_m(p)$
{} from  the  linear
behaviour   as   a   measure   of   intermittency.   It   holds
$\tilde\delta\zeta_m(p) \equiv\zeta_m(p)-m\zeta_3(p)/3=
\zeta_3(p)\delta\zeta_m(p)$.  In   the  VSR  it  is
$\zeta_3(p)\approx 1$, so both definitions for the intermittency corrections
essentially  agree,   but  in   the  VSR  we  find $\zeta_3(p)>1$,  so
$\tilde\delta\zeta_m(p)>\delta\zeta_m(p)$.}

The astonishing  results are shown in Fig.3a: There are large
intermittency corrections $\delta\zeta_m(p)$ for the small scales (large p,
VSR), only moderate intermittency corrections for the large
scales (small p, stirring subrange SSR), but hardly any  deviations
for p in the ISR.

The small  scale intermittency  is well  understood \cite{kra67,fri81}
and was  extensively discussed  in \cite{gnlo92b}. It is best seen in small
scale quantities as for example in the energy dissipation  rate
$\epsilon(\x,t)$
or in  the vorticity.  Here we  observe  in  addition  that  the
intermittency corrections  $\delta\zeta_m(p)$ in  the VSR  remarkably  well
agree with  the  r-scaling exponents  $\mu(m/3)$, defined by
$\langle\epsilon_r^{m/3}\rangle\propto
  r^{-\mu (m/3)}$,  which we  had already calculated in \cite{gnlo92b}.
This means
that for r in the VSR Kolmogorov's refined similarity hypothesis
(RSH) $v_r \propto (\epsilon_r r)^{1/3}$
seems to be well fulfilled. This also is in
agreement with Kraichnan's \cite{kra74b}, Frisch's \cite{fri91}, and
our \cite{gnlo93b} objections against the
RSH in  the ISR,  arguing that  for r  in the  ISR $v_r$  is an ISR
quantity, whereas  $\epsilon_r$ still  mainly is  a VSR  quantity. Thus  a
relation like  the RSH  should  only be expected, if r is in the
VSR and both $v_r$ and $\epsilon_r$ are VSR quantities.

Our result  is also  consistent with  latest full numerical
simulations \cite{hos92a},
which find  the RSH  fulfilled. Note  that in
these simulations  r is always in or at least near the VSR since
Re  is  still small.  And last not least our finding also agrees
with the  observation of  Chen et.al.\  \cite{che93} that the RSH is less
and less  fulfilled the larger r becomes. For further comparison
with experiment, see below.

Before we  interpret the  behaviour of  the $\delta\zeta_m(p)$  in the
stirring subrange  SSR and  in the  ISR, we checked how our
findings depend  on various changes of our Navier-Stokes
approximation: (i) To  be sure that the SSR-intermittency does not depend
on the  kind of  forcing (3),  we compared with  the alternative
forcing $\f(\p)\propto  \u(\p)$,  again $\p\in K_{in}$.
We also took a random forcing,
but the  results did  not change  noticeably. (ii) We varied the
set $K_{in}$  and allowed for more or for fewer modes which are stirred,
but  again there was no change. (iii) We varied the type of
wave vectors  in $K_0$  and their  number N  as well as the maximal
nonlocality $s_{max}$  of the  contributing triadic interactions (see
table  1).   Again,  no  sizeable  change.  In  particular,  the
intermittency  corrections  did  not  increase  with  increasing
nonlocality of the triadic interactions as we speculated in \cite{gnlo92b}.
(iv) Different values  of $\delta\zeta_ m(p)$  were only  obtained when  the
flow field  was not  yet statistically stationary, see Fig.4. In
Fig.4a we  averaged over  7 large  eddy turnover times only. The
total rate of  dissipated  energy   $\epsilon_{diss}=\nu\sum_{\p\in K}  \p^2
\langle |\u(\p)|^2\rangle$ still exceeds
the constant  input $\epsilon$   by about  $1\%$. In  this  case  there  are
considerable intermittency  corrections $\delta\zeta_m(p)$  for all p, which
go down  drastically in  the ISR  if one  averages over 70 large
eddy turnover  times, see  Fig.4b, where  we  had  statistically
stationary results. (Stationarity is identified from the balance
between the total dissipation rate and the total input rate.) Similar
observations have  been made when analysing experimental signals
\cite{vandewat91}.
(v) To  demonstrate how  $\delta\zeta_m(p)$ varies from run to run we
refer to Fig.5. The deviations $\delta\zeta_m(p)$ for p in the ISR are very
small, but  still seem  to be significant. (vi) We decreased the
degree of  locality of  the  $\zeta_m(p)$  by fitting  the larger range
$[p/\sqrt{20}, p\sqrt{20}]$,  see Fig.3b. Again no qualitative change;
$\delta\zeta_m(p)$
now  tends   to  become   even  smaller  in  the  ISR.  (vii) We
artificially extended  the ISR by putting $\nu =0$ and extracting the
energy from the smallest eddies by using a phenomenological eddy
viscosity as  employed in  \cite{egg91a}.
Now, as expected, $\delta\zeta_m(p)\approx 0$
also
for the  large p,  i.e., the  small scale  intermittency  really
originates from  the competition between transport downscale and
the     viscous  damping.   (viii) One  might   speculate   that
intermittency corrections in the ISR would occur if our
Fourier-Weierstrass ansatz  would not only be wave number but also space
resolving as  in \cite{egg91b}.  But when  doing this  we found  that the
intermittency corrections  observed in \cite{egg91b} vanish if the number
N of  wave vectors  in $K_0$ is increased \cite{gnlo92b}.
One should note that
we include  in fact  some degree  of position space localization
since any  Fourier representation with many modes already allows
for localization in space.

To have another check, we also
calculated the scale dependent
flatness $F(p)=\langle|\u(p)|^{4}\rangle/\langle|\u(p)|^{2}\rangle^{2}
\propto p^{-\zeta_4+2\zeta_2}$.
If there is intermittency, then $2\zeta_2 > \zeta_4$,
thus $F(p)$ has to increase with $p$. In fact we find
such an increase of $F(p)$ in the SSR from $F(p=3)\approx 2.7$
($< 3$, a result achieved also in various
full numerical simulations and
experiments, see e.g.\ \cite{she93} for a recent reference.)
to the value $F(p)\approx 3.0$
valid for a Gaussian distribution. For $p$ in
the ISR $F(p)=3$ stays constant. Approaching the VSR by further
increased p, the flatness now
strongly  grows \cite{gnlo92b}.
This can be understood as being due to the small scale
intermittency, as we extensively
reported in \cite{gnlo92b}.
This behaviour of $F(p)$ well agrees with the above described
findings for $\delta\zeta(p)$.

Finally we report how the flatness
of the velocity derivative $F_{1,1}=\langle (\partial_1u_1)^{4}
\rangle/\langle(\partial_1u_1)^{2}\rangle$
behaves as a function of $Re$. For the large
Reynolds numbers which we consider we find $F_{1,1}=3.15$,
independent of Re. This again means, we find
no intermittency as models which are
constructed to describe intermittency
obtain an increase of $F_{1,1}$ with the Re number in terms
of the ISR intermittency exponent $\mu
(2)$, $F_{1,1}\propto Re^{3\mu(2)/4}$, see e.g.
\cite{men87b,gnlo93b}.

Two conclusions of our findings are possible.

First, the very small if not missing
intermittency might be due to our approximation. Even
in the present, considerably improved ansatz the
larger wave vectors are still thinned
out, cf.Fig.1. If this argument was valid, the ISR-
intermittency would have been identified by
us as an effect of the opening of the phase space
for larger wave vectors. Consequently, there
should be no intermittency in 2D
turbulence, where the energy cascade is inverse -- and infact Smith and
Yakhot
\cite{smi93} do not find intermittency in numerical 2D turbulence.

The second possible conclusion is that
there indeed might be no intermittency in the pure ISR in the limit of large
Re.
Of course, if so that must be due to the
particular form  of the nonlinearity, namely the $\u\cdot grad \u$-term in the
Navier-Stokes equation.  It provides energy transport both
downscale and  upscale which,  as  our  solutions  show,  fluctuates
wildly and  with large amplitudes around a rather small mean
value of  downscale transport. This nearly symmetric down- and up-
scale transport is broken on the large scales (i.e., in the SSR)
due to  the finite size of the system. The largest eddies do not
get energy  by turbulent  transfer downscale  but  only  deliver
turbulent energy to smaller scales. The symmetry of transport is
also broken for small scales by the competition with the viscous
dissipation. May  be that the symmetry breaking mechanisms cause
the large and the small scale intermittency. Note that Galileian
invariance is only broken by the boundaries, i.e., by the finite
size of the system. Both dissipation and our forcing scheme keep
it.

This second possible conclusion is in
agreement with a recent theory developed by Castaing et.al.\
\cite{cas90}.
This theory predicts that $\delta\zeta_m=0$
in the limit of large Re, and that in this limit $F_{1,1}$
is independent of Re, which we find, as mentioned above.
The value of the $F_{1,1}$-limit, if it exists,
is probably larger than what we find.
Vincent and Meneguzzi \cite{vin91} calculate
already $F_{1,1}= 5.9$
for a Taylor Reynolds number $Re_\lambda\approx 100$.
Castaing et al.\ \cite{cas90}
find from their data analysis, that the flatness $F(p)$ increases as
$\log{F(p)}\propto (\eta p)^{\beta}$ with
$\beta\propto 1/ \log{(Re_\lambda/75)}$.
Our finding $F(p)\approx 3$ in the ISR
(for $Re\approx 10^{5}$, $Re_\lambda \approx  9000$)
well agrees with the large Re limit of this experimental behavior.
Note that in experiment it is still
$\beta\approx 0.24$ even for
$Re_\lambda= 2720$ \cite{cas90}.
Also the
measured intermittency
correction $\delta\zeta_m$ at
least for $m\ge 6$ are not 0 even for $Re_\lambda=2720$
\cite{cas90,ans84}. However
it could
well be that in experiment the intermittency
corrections might be overestimated,
because they will tend to increase if the averaging time is not
large enough and the flow is not yet
statistically stationary, see above, Fig.4, and the remarks in Ref.\
\cite{vandewat91}.

Our finding, that there might be three
ranges
for high Re turbulence -- namely,
the SSR with moderate intermittency, the ISR with practically no intermittency,
and the VSR with strong
intermittency -- might also
be supported by some experimental data arround.
In  Fig.6 the  spectrum $\langle |\u(\p)|^2\rangle$
is shown, taken from
Gagne's wind tunnel measurements
\cite{gag87} with the very high Reynolds number $Re_\lambda =2720$. While
in the  ISR  $\zeta_2=0.67$,  i.e., $\delta\zeta_2 = 0$,
seems to be a good fit, for
the large scales (SSR) the exponent  $\zeta_2=0.70$ is more appropriate.
In the VSR the exponential damping according to (4) is not
separated, so that  $\zeta_2$  cannot reliably be identified in that range.
Note, that in the same
experiment higher moments and the scale resolved flatness $F(p)$,
which is more sensitive to intermittency corrections, show
intermittency {\it also} in the ISR.

A similar  interpretation seems  possible by inspection of Praskovsky's
\cite{pra92} data  for $\langle|\v_r|^6\rangle$
measured at  the also very high $Re_\lambda =3200$, see Fig.7:
In the  middle of  the ISR we clearly have $\delta\zeta_6(p)=0$, whereas in
the VSR it takes the value $\delta\zeta_6=0.31$. This is precisely what one
expects from  the RSH, namely $\delta\zeta_6\approx\mu\approx 0.30$.
In  the SSR  the  intermittency
correction is  $\delta\zeta_6(p)=0.27$, which  is considerably  larger than
what we  found in this range. May  be this   is  due to the plumes, swirls, or
other structures  \cite{zoc90} which  detach from  the boundary  in real
flow and might increase the intermittency in the SSR.

Finally, we remark that also the
quasi-Lagrangian perturbation analysis of the Navier-Stokes equation, done by
Belinicher
and L'vov \cite{bel87}, leads to $\delta\zeta_m(p)=0$
for p in the ISR in the large Re limit.

To summarize, it can not yet be
ultimatively decided which of
the discussed conclusions of our
numerical data will turn out to be robust. Either there in fact
{\it is}
ISR-intermittency also for
$Re\to \infty$
as an effect of phase space opening for
large wave vectors (which by construction of
our approximation scheme we miss), or there is indeed no
ISR-intermittency in the limit of large Re \cite{cas90,bel87}.
To decide this alternative, it would be very helpful to at least allow
{\it some}
opening of phase space, e.g.,
to increase the number of wave vectors
per level as $\log k$ as already done in a 2D approximate solution
of the Navier Stokes equation \cite{vaz92}.
If then intermittency does not show up again,
we clearly have to favour the conclusion that there is no ISR-intermittency in
the large Re limit as our results demonstrate.

\newpage

\noindent
{\bf {\it Acknowledgements:}} We heartily thank Bernard Castaing for
very enlightening discussions and the referees for helpful comments.
D.\ L.\ thanks the Aspen Center of Physics for its hospitality. We
also
heartily thank  Itamar Procaccia and Reuven
Zeitak for  their hospitality  during our stays at the
Weizmann-Institute, Rehovot,  Israel.  Partial  support  by  the
German-Israel-Foundation (GIF)  is gratefully  acknowledged.  The  HLRZ
J\"ulich supplied us with computer time.

\newpage


\begin{thebibliography}{10}

\bibitem{bat49}
G.~K. Batchelor and A.~A. Townsend, Proc. R. Soc. London A {\bf 199},  238
  (1949).

\bibitem{kol41}
A.~N. Kolmogorov, CR. Acad. Sci. USSR. {\bf 30},  299  (1941).

\bibitem{wei48}
C.~F. von Weizs\"acker, Z. Phys. {\bf 124},  614  (1948);
W. Heisenberg, Z. Phys. {\bf 124},  628  (1948);
L. Onsager, Phys. Rev. {\bf 68},  286  (1945).

\bibitem{men91b}
Ch. Meneveau and K.~R. Sreenivasan, J. Fluid Mech. {\bf 224},  429  (1991).

\bibitem{gnlo92b}
S. Grossmann and D. Lohse, Z. Phys. B {\bf 89},  11  (1992);
Physica A {\bf 194},  519  (1993).

\bibitem{egg91a}
J. Eggers and S. Grossmann, Phys. Fluids A {\bf 3},  1958  (1991).

\bibitem{gnlo93g}
S. Grossmann and D. Lohse, Universality in turbulence, Preprint, Marburg, 1993.

\bibitem{sig77}
E.~D. Siggia, Phys. Rev. A {\bf 15},  1730  (1977);
R.~M. Kerr and E.~D. Siggia, J. Stat. Phys. {\bf 19},  543  (1978);
M.~H. Jensen, G. Paladin, and A. Vulpiani, Phys. Rev. A {\bf 43},  798  (1991).

\bibitem{foi90}
C. Foias, O. Manley, and L. Sirovich, Phys. Fluids A {\bf 2},  464  (1990).

\bibitem{my75}
A.~S. Monin and A.~M. Yaglom, {\em Statistical Fluid Mechanics} (The MIT Press,
  Cambridge, Massachusetts, 1975).

\bibitem{eff87}
H. Effinger and S. Grossmann, Z. Phys. B {\bf 66},  289  (1987).

\bibitem{gnlo93a}
S. Grossmann and D. Lohse, Phys. Lett. A {\bf 173},  58  (1993).

\bibitem{she93b}
Z.~S. She and E. Jackson, Phys. Fluids A {\bf 5},  1526  (1993).

\bibitem{she93}
Z.~S. She {\it et~al.}, Phys. Rev. Lett. {\bf 70},  3251  (1993).

\bibitem{kra67}
R. Kraichnan, Phys. Fluids {\bf 10},  2080  (1967).

\bibitem{fri81}
U. Frisch and R. Morf, Phys. Rev. A {\bf 23},  2673  (1981).

\bibitem{kra74b}
R. Kraichnan, J. Fluid Mech. {\bf 62},  305  (1974).

\bibitem{fri91}
U. Frisch, Proc. R. Soc. London A {\bf 434},  89  (1991).

\bibitem{gnlo93b}
S. Grossmann and D. Lohse, Europhys. Lett. {\bf 21},  201  (1993).

\bibitem{hos92a}
I. Hosokawa and K. Yamamoto, Phys. Fluids A {\bf 4},  457  (1992);
J. Phys. Soc. Japan. {\bf 60},  1852  (1991) and addendum
J. Phys. Soc. Japan. {\bf 62},  380  (1993);
J. Phys. Soc. Japan. {\bf 62},  10  (1993).

\bibitem{che93}
S. Chen, G.~D. Doolen, R.~H. Kraichnan, and Z.~S. She, Phys. Fluids A {\bf 5},
  458  (1993).

\bibitem{vandewat91}
W. van~de Water, B. van~der Vorst, and E. van~de Wetering, Europhys. Lett. {\bf
  16},  443  (1991).

\bibitem{egg91b}
J. Eggers and S. Grossmann, Phys. Lett. A {\bf 156},  444  (1991).

\bibitem{men87b}
Ch. Meneveau and K.~R. Sreenivasan, Nucl. Phys. B (Proc. Suppl) {\bf 2},  49
  (1987);
M. Nelkin, Phys. Rev. A {\bf 42},  7226  (1990);
J. Eggers and S. Grossmann, Phys. Lett. A {\bf 153},  12  (1991).

\bibitem{smi93}
L.~M. Smith and V. Yakhot, Phys. Rev. Lett. {\bf 71},  352  (1993).

\bibitem{cas90}
B. Castaing, Y. Gagne, and E.~J. Hopfinger, Physica D {\bf 46},  177  (1990);
B. Castaing, J. Phys. {\bf 50},  147  (1989);
B. Castaing, Y. Gagne, and M. Marchand, Physica D  (1993), to appear.

\bibitem{vin91}
A. Vincent and M. Meneguzzi, J. Fluid Mech. {\bf 225},  1  (1991).

\bibitem{ans84}
F. Anselmet, Y. Gagne, E.~J. Hopfinger, and R.A. Antonia, J. Fluid Mech. {\bf
  140},  63  (1984).

\bibitem{gag87}
Y. Gagne, Thesis, University of Grenoble, 1987.

\bibitem{pra92}
A.~A. Praskovsky, Phys. Fluids A {\bf 4},  2589  (1992).

\bibitem{zoc90}
G. Zocchi, E. Moses, and A. Libchaber, Physica A {\bf 166},  387  (1990).

\bibitem{bel87}
V.~I. Belinicher and V.~S. L'vov, Sov. Phys. JETP {\bf 66},  303  (1987).

\bibitem{vaz92}
E. Vazquez-Semadeni and J.~H. Scalo, Phys. Fluids A {\bf 4},  2833  (1992).

\bibitem{kol62}
A.~N. Kolmogorov, J. Fluid Mech. {\bf 13},  82  (1962).

\bibitem{fri78}
U. Frisch, P.~L. Sulem, and M. Nelkin, J. Fluid Mech. {\bf 87},  719  (1978).

\bibitem{sre93}
K.~R. Sreenivasan and P. Kailasnath, Phys. Fluids A {\bf 5},  512  (1993).

\end{thebibliography}

\newpage
\centerline{{\bf Table 1}}
 \begin{table}[htp]
 \begin{center}
 \begin{tabular}{|l|l|l|l|}
 \hline
       & interaction
       &
       &
 \\
       & number of
       &
       &
 \\
          N
       & triades
       & $s_{max}$
       & $b$
 \\
 \hline
        26 \cite{egg91a}
       & 39
       & 1.92
       & 300
 \\
        38 \cite{gnlo92b,egg91a}
       & 102
       & 1.92
       & 170
 \\
        50
       & 273
       & 3.46
       & 80
 \\
        74
       & 741
       & 5.00
       & 70
 \\
        74
       & 729
       & 5.74
       & 70
 \\
        74
       & 858
       & 3.46
       & 70
 \\
        80
       & 783
       & 5.74
       & 70
 \\
        86
       & 966
       & 5.74
       & 65
 \\
  \hline
 \end{tabular}
 \end{center}
 \end{table}

\vspace{1cm}
\noindent
Characteristic parameters  of several different wavesets $K_0$. N
denotes the  number of wave vectors in $K_0$. In the second column the
number of  triadic interactions  $\p=\q_1+\q_2$ between the wave vectors of
any one  set $K_l$ is given, $l$  not too large or small to avoid edge effects.
$s_{max}$ is the maximal nonlocality (definition see text)
of the  triadic interactions. b is the dimensionless constant in
the  structure   function  $D(r)=\langle |\u(\x+\r)-\u(\x)|^2\rangle
=b(\epsilon r)^{2/3}$, $r\in$ ISR.   The
experimental value  is b=8.4  \cite{my75}.  The  larger  values  in  our
approximation can well be understood \cite{egg91a}, the decrease of b with
the increase of N is in keeping with that explanation.

\newpage
\centerline{{\bf Table 2}}
 \begin{table}[htp]
 \begin{center}
 \begin{tabular}{|l|l|l|l|l|l|l|}
 \hline
       &  $\langle |\u |^2 \rangle$
       &  $\langle |\u |^3 \rangle$
       &  $\langle |\u |^4 \rangle$
       &  $\langle |\u |^6 \rangle$
       &  $\langle |\u |^8 \rangle$
       &  $\langle |\u |^{10} \rangle$
 \\
 \hline
        $\zeta_m$
       & 0.682
       & 1.021
       & 1.359
       & 2.034
       & 2.707
       & 3.370
 \\
        $\zeta_m/\zeta_3$
       & 0.668
       & 1.000
       & 1.331
       & 1.993
       & 2.651
       & 3.301
 \\
        $\delta\zeta_m$
       & 0.002
       & 0.000
       & 0.002
       & -0.007
       & -0.016
       & -0.033
 \\
        $\delta\zeta_m(K62)$
       & 0.022
       & 0.000
       & -0.044
       & -0.200
       & -0.444
       & -0.778
 \\
        $\delta\zeta_m(ran-\beta)$
       & 0.038
       & 0.000
       & -0.046
       & -0.170
       & -0.347
       & -0.590
 \\
        $p_{D,m}$
       & 597
       & 398
       & 299
       & 201
       & 152
       & 122
 \\
        $2p_{D,2}/m$
       & 597
       & 398
       & 299
       & 199
       & 149
       & 119
 \\
  \hline
 \end{tabular}
 \end{center}
 \end{table}

\vspace{1cm}
\noindent
Results from  the fit (4) to the spectra $\langle|\u(\p)|^m\rangle$ obtained
with N=86  wave vectors in  $K_0$ for  moments up to m=10. $\nu =8\cdot
10^{-6}$.
The average  is over  80 large eddy turnover times (skipping the
first 100 turnovers). We fitted the p-range $[0,1000]$. In general
$\zeta_3$ is rather near, but not exactly 1 as it should be according
to Kolmogorov's  structure equation  \cite{my75}. We  therefore calculate
the  intermittency   corrections  from   renormalized  exponents
$\zeta_m/\zeta_3$, namely, $\delta\zeta_m= \zeta_m/\zeta_3-m/3$.
For comparison, the values for
Kolmogorov's log  normal model  \cite{kol62},
$\delta\zeta_m=-\mu m(m-3)/18$, are also
given, which are well known to fit the
data for $m\le 10$ with $\mu = \mu_2 \approx
0.20$.
$\delta\zeta_m(ran-\beta)=-\log_2[1-x+x(1/2)^{1-m/3}]$
are     the    intermittency
corrections due  to the  random $\beta$-model  (x=0.125) \cite{fri78}.
$p_{D,m}$ is
the dissipative  cutoff, which  agrees very well with $2p_{D,2}/m$ as
shown in the last row.

\newpage
\centerline{{\bf Figure captions}}

\vspace{0.3cm}\noindent
\underline{Figure 1}\\
Comparison of  the  subset  of  wave  numbers  admitted  in  the
Fourier-Weierstrass decomposition  (upper) with  the complete p-
spectrum (lower),  either $0\le |\p|\le 25$  or $0\le |\p| \le 500$.
While the exact
density of  wave numbers  increases as $p^2$, the geometric scaling
$2^l K_0$ makes  the density  decrease, i.e.,  the smaller scales are
less well resolved.

\vspace{0.3cm}\noindent
\underline{Figure 2}\\
Spectra $\langle |\u(\p)|^m\rangle$ for m=2 ($\diamond$) and m=6 (+).\\
\noindent
a) N=86 wave vectors in $K_0$, $\nu =8\cdot 10^{-6}$, $s_{max}=5.74$,
$Re_\lambda =9030$, averaging time  from 100 to 180.
The input set $K_{in}$ of the forcing (3) consists of the 12 shortest
wave vectors. The lines are the fits (4).
\\ \noindent
b) N=38, $\nu =8\cdot 10^{-6}$, averaging time 150 large eddy turnovers, same
forcing as in (a).

\vspace{0.3cm}\noindent
\underline{Figure 3}\\
$\delta\zeta_m(p)$ for m=2,4,6,8,10, bottom to top. Same data as in Fig.2a.
The shaded  ranges on  the right show the Kolmogorov values \cite{kol62}
$\delta\zeta_m=-\mu m(m-3)/18$ for  $\mu=0.20$ through  $\mu=0.30$,
because  in   \cite{sre93}
$\mu=0.25\pm 0.05$ is given as "best estimate". In (a) the fit range is
$[p/\sqrt{10},
p\sqrt{10}]$, in (b) the larger local range $[p/\sqrt{20}, p\sqrt{20}]$ is
chosen.

\vspace{0.3cm}\noindent
\underline{Figure 4}\\
$\delta\zeta_m(p)$  for  m=2,4,6,8,10  (bottom  to  top).  N=80,
$\nu=5\cdot 10^{-6}$,
$s_{max}=5.74$, $Re_\lambda=13~550$.
The averaging times are  (a) 7 and (b) 70 large
eddy turnovers, respectively.

\vspace{0.3cm}\noindent
\underline{Figure 5}\\
(a) $\delta\zeta_6(p)$   and (b) $\delta\zeta_2(p)$ for  N=80,
$\nu =5\cdot 10^{-6}$,  $s_{max}=5.74$ for
different runs.  For two runs the averaging time is about 30 large
eddy turnovers,  for another  two  runs  it  is  about  70.  The
(weighed) means  and the  standard deviations are marked by a dia-
mond and by error bars, respectively.

\vspace{0.3cm}\noindent
\underline{Figure 6}\\
Spectrum $E(r/\eta )$  from [16]  for $Re_\lambda =2720$.
Three ranges  can  be
identified. On  the large  scales the SSR with   $\zeta_2+1=1.70$, i.e.,
some intermittency  corrections, for  moderate r  the  ISR  with
$\zeta_2+1=5/3$, i.e., no intermittency, and for small r the VSR.

\vspace{0.3cm}\noindent
\underline{Figure 7}\\
Moments  of   the  velocity   differences
$\langle |v_r|^6\rangle /r^2$ ($\circ$)
and   of  the  energy
dissipation rate $\langle\epsilon_r^2\rangle$  ($\bullet$)
(arb.units) against r, taken
{} from [27].  The slopes of these quantities are  $\delta\zeta_6(p)$
and $\mu=\mu_2$,
respectively. Clearly,  for $\langle |v_r|^6\rangle /r^2$
there are three ranges VSR,
ISR, and  SSR, whereas  the VSR  quantity  $\langle\epsilon_r^2\rangle$
does  not  show
different ranges.  The  arrows  correspond  to  $20\eta /L $ and  $1/5$,
respectively.

\end{document}